\documentstyle[aps,twocolumn,graphicx,epsfig]{revtex}
\newcommand{\be}{\begin{equation}}
\newcommand{\ee}{\end{equation}}

\begin{document}

\draft

\widetext
\title{Dynamics of a Bose-Einstein condensate at finite temperature in an atomoptical
coherence filter}

\author {F.~Ferlaino, P.~Maddaloni,\cite{pm} S.~Burger, F.~S.~Cataliotti, C.~Fort,
M.~Modugno, and M.~Inguscio}

\address{INFM--European Laboratory for Nonlinear Spectroscopy (LENS) and Dipartimento di Fisica, Universit\`a di Firenze,\\L.go E. Fermi 2, I-50125 Firenze, Italy}

\date{\today}

\maketitle
\begin{abstract}

The macroscopic coherent tunneling through the barriers of a
periodic potential is used as an atomoptical filter to separate
the condensate and the thermal components of a $^{87}$Rb mixed
cloud. We condense in the combined potential of a laser
standing-wave superimposed on the axis of a cigar-shape magnetic
trap and induce condensate dipole oscillation in the presence of
a static thermal component. The oscillation is damped due to
interaction with the thermal fraction and we investigate the role
played by the periodic potential in the damping process.

\end{abstract}

 \narrowtext

\narrowtext

The combination of Bose-Einstein condensates (BECs) with periodic
optical potentials provides a versatile tool for getting a deeper
understanding of macroscopic quantum phenomena, directly
addressing the issue of phase coherence between degenerate coupled
Bose gases \cite{kasevich98,kasevich,hansch,pedri} recently
culminating in the observation of the Mott insulator phase
transition \cite{bloch}. Properties of condensates confined in an
optical lattice have opened new perspectives in the investigation
of different dynamical phenomena such as superfluid flow
\cite{sven}, atomic Bloch oscillations \cite{arimondo}, the
Josephson effect \cite{science} and have stimulated speculations
on a wide range of phenomena including solitons \cite{solitoni},
dynamical instability \cite{instabilita} and transport behavior
\cite{trasporto}.

The dynamics of BECs at finite temperatures, is a challenging
question even in the pure harmonic case. Many efforts have been
devoted to the extension of theoretical models at nonzero
temperature and to the study of the low-energy collective
excitations in presence of a significant thermal component
\cite{zaremba,giorgini,griffin,shlap1,ZGN,jackson}.

From the experimental point of view the effects of the
interactions between the condensate and the thermal fraction on
the modes of excitation have been investigated only in harmonic
potentials. Temperature dependence of damping and frequency
shifts in quadrupole and scissors modes have been investigated in
a TOP trap \cite{cornell,foot} while also a damped out-of-phase
dipolar oscillation of the thermal cloud and the condensate has
been observed in a cloverleaf trap \cite{ketterle}.

In this Letter we address the problem of the dynamics of a
condensate in a periodic potential at finite temperature. This is
done by using an optical lattice as an atomoptical filter to
induce, through a selective manipulation, a relative motion
between the condensate and thermal components of a magnetically
trapped mixed cloud of rubidium atoms. In the presence of the
periodic potential these two components respond differently to a
sudden displacement of the magnetic trap. Atoms with sufficiently
long coherence lengths are allowed to tunnel through the potential
barriers of a one dimensional optical lattice while atoms with a
higher kinetic energy remain confined in the potential wells.
While the condensate coherently flows through the optical
barriers, thus performing a dipole oscillation, the thermal
component is blocked in the presence of the periodic potential and
provides a damping mechanism. In particular, we observe a sharp
change of the damping rate in the BEC dipole oscillations by
increasing the optical potential depth.

Our procedure for creating Bose condensates in an optical lattice,
already described in details elsewhere \cite{sven}, can be
summarized as follows. Starting with a sample of ultracold atoms
of $^{87}$Rb confined in a Ioffe-type magnetic trap in the
$(F=1,m_{F}=-1)$ state, we perform rf-evaporative cooling until we
reach a temperature slightly above the condensation threshold.
Next, we superimpose a one-dimensional optical lattice to the
longitudinal $x$-axis of the magnetic trap by means of a far
detuned, retroreflected laser beam. We then continue the
evaporation process through the phase-transition temperature
\cite{nota}. In this way, the atomic cloud is prepared in an
equilibrium state of the combined magnetic and optical dipole
potentials. In this experiment we typically stop the evaporation
when the fraction of condensed atoms is $\simeq 20\%$
corresponding to a temperature $T \simeq 120$~nK. In the atomic cloud region the
resulting potential has the form
\begin{eqnarray}
V &=&V_{mag}+V_{opt}\nonumber \\
&=&\frac{1}{2}m [ \omega _{x}^{2}x^{2}+\omega _{\perp }^{2}(
x^{2}+y^{2}) ] +sE_{R}\cos ^{2}\left( \frac{ 2\pi x}{\lambda
}\right) \label{pote}
\end{eqnarray}
with $m$ the atomic mass, $\omega _{x}=2\pi \times 9$~Hz and
$\omega _{\perp }=2\pi \times 92$~Hz  the axial and radial
frequency of the magnetic potential respectively. In (\ref{pote})
the optical potential depth is expressed, via the dimensionless
factor $s$, in units of the recoil energy of an atom absorbing one
laser photon, $E_{R}=h^{2}/2m\lambda ^{2}$, where $\lambda $ is
the wavelength of the laser creating the standing wave.

By varying the intensity of the laser beam, (detuned typically
$\Delta\lambda \simeq 3$~nm to the blue of the $D_{1}$ line at
$\lambda \simeq 795$~nm), we change the value of $s$ up to 2.5.
The optical potential is then calibrated by measuring the Rabi
frequency of the Bragg transition between the momentum states
$-h/\lambda $ and $+h/\lambda$ induced by the standing wave
\cite{salo}. Due to the large detuning of the optical lattice,
the maximum spontaneous scattering rate is $\Gamma_{sp}=
0.25$~Hz and spontaneous scattering can be
neglected on the timescale of our experiment.

In order to induce the relative center-of-mass motion of the
condensate and the thermal fractions in the combined trap, we
first displace instantaneously ($t_{dis}\ll 2\pi/ \omega _{x}$)
the magnetic trapping potential in the $ x$-direction by a
distance $\Delta x = 18 \mu$m sufficiently small to remain in the
superfluid regime \cite{sven}. After the displacement, the atomic
cloud finds itself out of equilibrium, nevertheless due to the
presence of the optical lattice, only the condensate is forced
into motion by the magnetic potential gradient. Indeed, when the
potential energy given by the displacement is smaller than the
optical potential depth, as in our experiment, a motion along the
$x$-direction can be established only by means of tunneling
through the barriers of the standing wave. We have previously
shown \cite{science} that a pure condensate and a thermal cloud
above T$_c$ behave in a dramatically different way. The coherence
properties of the condensate imply a macroscopic tunneling and a
collective dipole motion can occur. On the contrary, the
incoherent nature of the thermal cloud drastically reduces the
collective tunneling probability, thus preventing the
establishing of a collective oscillation. In the present work we
produce mixed clouds and the optical lattice acts as a filter to
separate the coherent and incoherent components. After the
displacement, we wait half a period of the BEC oscillation in the
harmonic trap ($\pi /$ $\omega _{x}$) then we move back the
magnetic potential by the distance $-\Delta x$. At the end of
this procedure, the thermal cloud finds itself at rest in the
center of the trap, while the condensate oscillates inside the
thermal cloud with an initial amplitude $2\Delta x$. After a
variable evolution time, the trapping potential is switched off
and the cloud is imaged subsequently to an additional ballistic
expansion of 28~ms. From the observed column density we deduce
the condensate fraction, the temperature and the center-of-mass
position of both the condensate and the thermal components in the
trap. In Fig.~1 we report three absorption images corresponding
to successive times of the evolution in the combined trap.

In order to show the effects of the optical lattice on the
dynamics of a mixed cloud, we study the center-of-mass motion of
the condensate and the thermal components in two different
configurations. In one ({\em harmonic case}) we switch off the
periodic potential after the preparatory procedure described
above while in the other ({\em combined case}) we leave the
lattice on also during the subsequent evolution.

{\em Harmonic case}~-~In absence of the lattice both the
condensate and the thermal cloud are free to oscillate. Starting
from an off-equilibrium configuration in which the thermal cloud
is at the center of the trap, the condensate mean-field pushes the
thermal fraction out of the potential minimum. During the time
evolution the oscillation amplitude of the thermal cloud rises
until an equilibrium is reached (Fig.~2a). Notice that, in a pure
harmonic trap, when no relative motion is induced, the BEC
dipolar mode can not be affected by the presence of the thermal
component \cite{revmod}. We expect that the kinetic energy lost
by the condensate is essentially converted into kinetic energy of
the thermal cloud. From the fit of the relative center-of-mass
position (Fig.~2b) with a damped oscillation, we obtain a
frequency of $8.42 \pm 0.04$~Hz which is smaller of roughly $5\%$
from the measured trapping frequency. This is a consequence of
the interaction between the oscillating condensate and the
initially stationary thermal cloud and is similar to the
observation reported in \cite{ketterle} for an oscillating
thermal cloud in presence of an initially stationary condensate.

{\em Combined case}~-~In presence of the optical lattice the
thermal cloud stays locked at the center of the trap due to its
incoherent nature, while the condensate oscillates as shown in
Fig.~3. Again the coupling between the condensate ad the thermal
cloud causes damping in the center-of-mass oscillations of the
condensate. Here, the kinetic energy lost by the BEC during the
evolution cannot be converted into the kinetic energy of the
thermal cloud which is kept fixed by the periodic potential and
must be therefore partly converted into internal energy of the
cloud, partly absorbed by the optical lattice. Note that in this
case the pure condensate would perform undamped oscillations
\cite{sven}.

We investigate the damping rate of the condensate center-of-mass
motion as a function of the optical potential depth $s E_R$ for a
fixed trap displacement. The experiment is performed in the
collisionless regime where collisions between thermal atoms are
negligible. In this regime Landau damping, present when a thermal
bath of elementary excitations absorbs quanta of the condensate
collective excitations, represents an important mechanism to
explain the dynamical behaviour of trapped Bose gases
\cite{Liu97}.

For temperatures larger than the chemical potential, the Landau
damping rate of low-energy excitations in a harmonically trapped
Bose-condensed gas can be estimated using the form \cite{revmod}
\be \label{gamma} \Gamma_L=\frac{3 \pi}{8} \frac{k_BTa
\omega}{\hbar c} \ee where $c=\hbar \sqrt{4\pi a n_0}/m$ is the
sound velocity in the center of the trap, $n_0$ being the central
density of the condensate, $a$ the {\em s-}wave scattering length
and $\omega$ the frequency of the collective excitation. In our
case, for small values of $s$, we can estimate $\Gamma_L$
calculating the central density of the condensate in presence of
the periodic potential using a variational approach with a
gaussian ansatz for the condensate wave-function. For $s=0.7$ and
assuming a fixed temperature $T=120$~nK, we have $\Gamma_L
=1.6$~Hz in fair agreement with the experimental results.

In Fig.~4 we show the damping rate measured as a function of the
optical potential depth. Close to $s=1$ we observe a sudden
increase of the damping rate. This occurs in coincidence with the
formation of the first bound state of the condensate in the
lattice \cite{molmer}. Actually the occurrence of a band
structure strongly modifies the energy spectrum. In particular
the Landau damping, being a resonant energy transfer, strongly
depends on the form of energy spectrum \cite{shlap1} and an
adequate treatment would be needed. In addition a theoretical
analysis of finite temperature damping of excitations in presence
of a periodic potential should also include other damping
mechanisms such as inter-component damping recently suggested in
\cite{griffin}. However, at present, a comprehensive theoretical
analysis which accounts for all the effects introduced by the
presence of the lattice is lacking.

In conclusion, we have demonstrated the operation of an
atomoptical filter capable of separating the condensate and the
thermal fraction in a mixed atomic cloud across T$_c$. This
filtering technique could, e.~g., be applied to spatially
separate the ground state of a finite temperature BEC fully from
the thermal component. In this work, we are able to control the
relative motion between the center-of-masses of the condensate
and of the thermal fraction in a mixed cloud, this enable us to
investigate quantitative effects both in the amplitude and in the
frequency of the dipole oscillations. Damping is caused by
interactions between condensate and the thermal fractions. Novel
features are expected in presence of the periodic lattice
potential which modifies the energy spectrum of the system.
Indeed by changing the optical lattice parameters a sharp rise in
the damping rate has been observed when a bound state is likely
to be formed.

We are indebted to Francesco Minardi for many useful discussions
and for his help in the initial stage of the experiment. We
acknowledge Gabriele Ferrari, Sandro Stringari, Stefano Giorgini
and Lev Pitaevskii for fruitful discussions, and Maurizio Artoni
for careful reading of the manuscript.

This work has been supported by the EU under Contracts No.
HPRI-CT 1999-00111 and No. HPRN-CT-2000-00125, by the MURST
through the PRIN 1999 and 2001 Initiatives and by the INFM
Progetto di Ricerca Avanzata ``Photon Matter''.

\begin{figure}
\label{foto}
\begin{center}
\includegraphics[width=8cm]{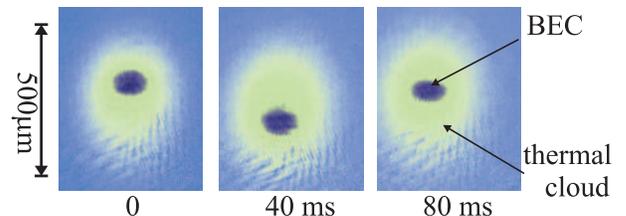}
\caption{Absorption images after different evolution times in the
combined trap showing the dipole oscillations of the condensate
inside the thermal cloud.}
\end{center}
\end{figure}

\begin{figure}
\begin{center}
\includegraphics[width=8cm]{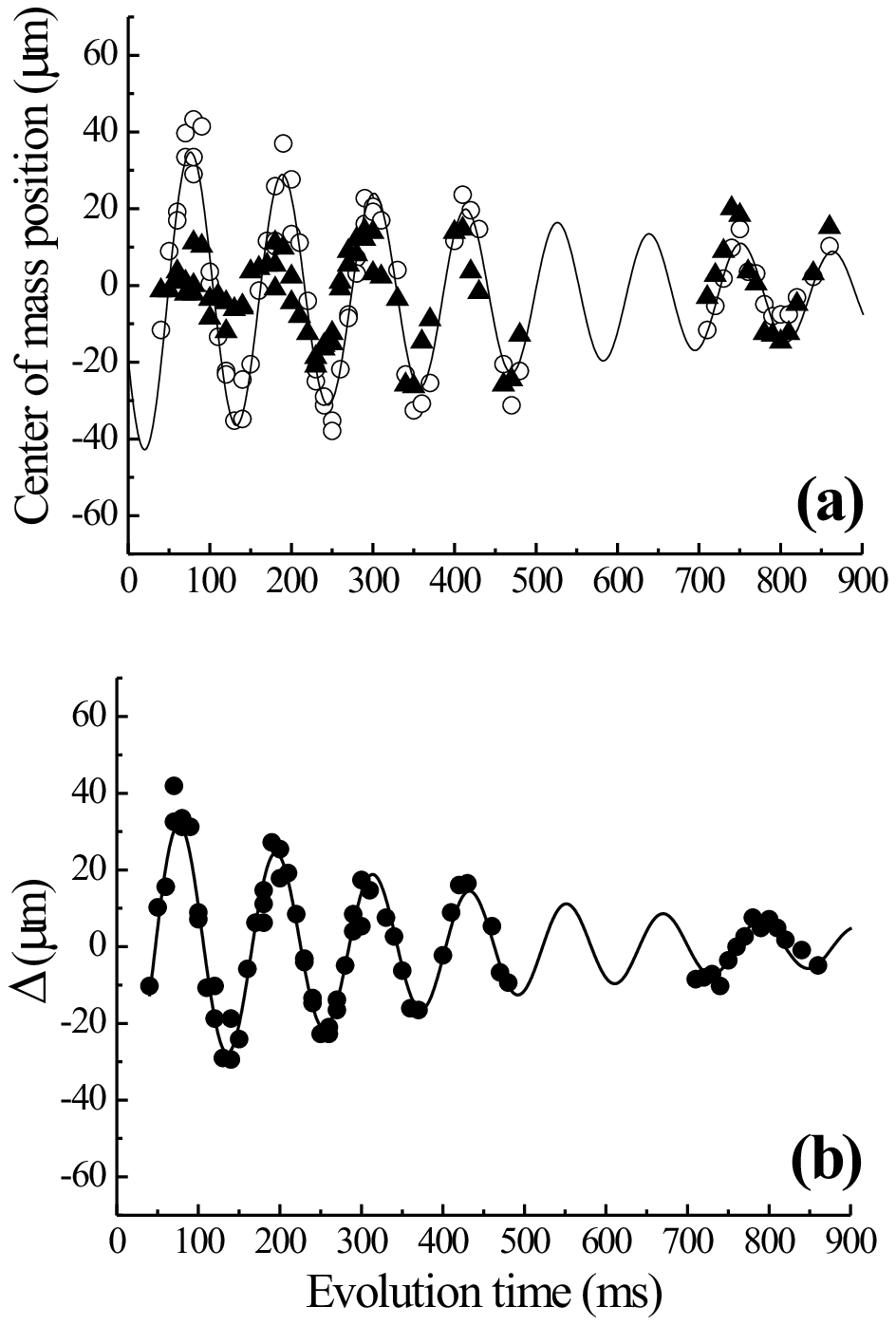}
\caption{a) Center of mass position of the condensate (open
circles) and the thermal cloud (filled triangles) after the
expansion as a function of the time spent in trap when the optical
potential is switched off after the initial displacement (the
line is a guide to the eye). b) Difference $\Delta$ between the
two components center-of-mass position plotted together with a
fit to the data (continuous line).}
\end{center}
\end{figure}

\begin{figure}
\begin{center}
\includegraphics[width=8cm]{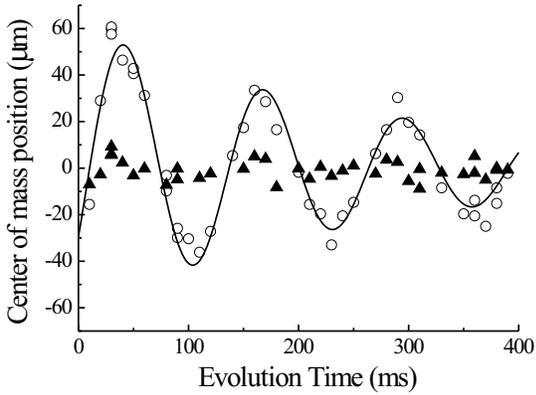}
\caption{Center-of-mass position of the condensate (open circles)
and the thermal cloud (filled triangles) after the expansion as a
function of the time spent in the combined trap ($s=1.8$)
together with a fit to the condensate center-off-mass position
(continuous line).}
\end{center}
\end{figure}

\begin{figure}
\begin{center}
\includegraphics[width=8cm]{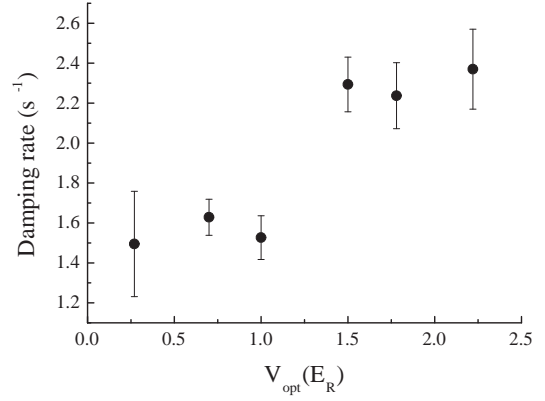}
\caption{Damping rate $\Gamma$ as a function of the optical
potential depth in units of the recoil energy. The occurrence of a
band structure ($s>1$) is reflected in a jump of the damping
rate.}
\end{center}
\end{figure}

\end{document}